\documentclass[USenglish]{article}	

\usepackage[utf8]{inputenc}				
\usepackage[big,online]{dgruyter}	
\usepackage{lmodern} 
\usepackage{microtype}
\usepackage[numbers,square,sort&compress]{natbib}

\theoremstyle{dgthm}

\theoremstyle{dgdef}

\begin{document}

	\articletype{Research Article}
	\received{July	6, 2021}
  \accepted{August	27, 2021}
  \journalname{Nanophotonics}
  \journalyear{2021}
  \journalvolume{XX}
  \journalissue{X}
  \startpage{1}
  \aop
  \DOI{10.1515/nanoph-2021-0351}

\title{Bound states in the continuum in strong-coupling and weak-coupling regimes under the cylinder - ring transition}
\runningtitle{BIC in strong and weak-coupling regimes}

\author[1]{Nikolay Solodovchenko}
\author[2]{Kirill Samusev} 
\author[1]{Daria Bochek}
\author*[3]{Mikhail Limonov} 
\affil[1]{\protect\raggedright 
ITMO University, Department of Physics and Engineering, St. Petersburg 197101, Russia, e-mail:  n.solodovechenko@metalab.ifmo.ru, dashabocheck@gmail.com}
\affil[2]{\protect\raggedright 
ITMO University, Department of Physics and Engineering, St. Petersburg 197101, and Ioffe Institute, St. Petersburg 194021, Russia, e-mail: k.samusev@mail.ioffe.ru}
\affil[3]{\protect\raggedright 
ITMO University, Department of Physics and Engineering, St. Petersburg 197101, and Ioffe Institute, St. Petersburg 194021, Russia, e-mail:  m.limonov@mail.ioffe.ru}
	
	
\abstract{Bound states in the continuum (BIC) have been at the forefront of research in optics and photonics over the past decade. It is of great interest to study the effects associated with quasi-BICs in the simplest structures, where quasi-BICs are very pronounced. An example is a dielectric cylinder, and in a number of works, quasi-BICs have been studied both in single cylinders and in structures composed of cylinders. In this work, we studied the properties of quasi-BICs during the transition from a homogeneous dielectric cylinder in an air environment to a ring with narrow walls while increasing the diameter of the inner air cylinder gradually. The results demonstrate the quasi-BIC crossover from the strong-coupling to the weak-coupling regime, which manifests itself in the transition from avoided crossing of branches to their intersection with the quasi-BIC being preserved on only one straight branch. In the regime of strong-coupling and quasi-BIC, three waves interfere in the far-field zone: two waves corresponding to the resonant modes of the structure and the wave scattered by the structure as a whole. The validity of the Fano resonance concept is discussed, since it describes the interference of only two waves under weak coupling conditions.}


\maketitle

\section{Introduction} 
Localization of electromagnetic waves is important both for fundamental research and for many important applications. Various physical mechanisms of trapping and confining of light are exploited including BICs, which have been actively investigated recently. BIC is a general wave phenomenon that was first mathematically proposed in 1929 by von Neumann and Wigner \cite{Wigner:1929} for electronic states. Note that the initial proposal of von Neumann and Wigner was never implemented in practice, but the idea of the appearance of such a state turned out to be very fruitful; a number of other mechanisms of the BIC formation were theoretically proposed and experimentally demonstrated \cite{Marinica:2008, Bulgakov:2008, Plotnik:2011, Hsu:2013, Yang:2014, Monticone:2014, Hsu:2016, Rybin:2017, Doeleman:2018, Monticone:2018, Azzam:2018, Koshelev:2019, Krasnok:2018, Bogdanov:2019, Koshelev:2020}. Different BICs has been observed for various types of waves such as electromagnetic, acoustic, elastic and water waves. 

Photonic BICs coexist with propagating electromagnetic waves and lie in a continuum, but theoretically remain completely confined in the structure without any radiation \cite{Hsu:2016}. According to the theory, BICs are unique states with an infinite lifetime and can arise if at least one dimension of the structure extends to infinity \cite{Hsu:2016} or in structures of finite length, when the dielectric constant approaches zero \cite{Monticone:2014,Silveirinha:2014}. In reality, due to the finite length of the structures, material loss, and imperfection, the BICs collapse to a quasi-BIC with a limited radiation $Q$-factor \cite{Azzam:2021}. The quasi-BIC is achieved when the Q-factor no longer changes slowly as in a conventional resonator, but instead rises rapidly, following the BIC trend, before it reaches its maximum value, due to the finite size of the resonator \cite{Rybin-Nat:2017}. In accordance with the classification presented in Review \cite{Hsu:2016}, three types of BIC (or quasi-BIC) can be distinguished. First, there are BICs protected by symmetry and separability; second, these are BICs built using inverse construction (for example, potential, hopping rate, or shape engineering); and finally, there are BICs, achieved by setting parameters. 

In this work, we will be interested in the third type of quasi-BIC. This type of quasi-BIC occurs when two non-orthogonal modes are coupled to the same radiation channel and a strong-coupling regime of avoided crossing arises with appropriate conditions in parametric space \cite{Azzam:2018,Doeleman:2018,Bulgakov:2018,Odit:2020,Lee:2020,Seo:2020}. This regime is described by the Friedrich-Wintgen model \cite{Friedrich:1985} when due to destructive interference one of the emitting modes disappears and the other becomes more lossy due to constructive interference. In photonic crystal slabs, such a regime was described by Tikhodeev et al \cite{Tikhodeev:2002} and Christ et al \cite{Christ:2003}. Recently Friedrich-Wintgen's quasi-BIC was observed and studied in detail in dielectric cylinders \cite{Rybin:2017,Bogdanov:2019}. For a single dielectric cylinder, quasi-BIC occurs when two eigenmodes with different polarizations, associated with the Mie resonances and Fabry-Perot resonances, in the strong-coupling regime form an avoided crossing region. These modes are approximately orthogonal inside the cylinder and they interfere predominantly outside, realizing the quasi-BIC when the all-dielectric resonator demonstrates extremely high values of the $Q$-factor. 

Here we continue the search and study of quasi-BIC in the simplest dielectric objects, which are bodies of revolution \cite{Rybin:2017,Bogdanov:2019}. We are looking for an answer to the question - what happens with the quasi-BIC during a significant transformation of the structure which even changes its topology? We analyze the transformation of quasi-BIC when adding a coaxial air hole to a cylinder with a high refractive index and gradually increasing the radius of the hole, that is, turning the cylinder into a narrow ring. As it often so happens, deviations from perfect structure can result in higher complexity and give rise to unexpected effects. Indeed, we discovered and investigated in detail the quasi-BIC crossover from the strong-coupling to the weak-coupling regime, which manifests itself in the transition from avoided crossing of branches to their intersection with the quasi-BIC being preserved on only one branch. We also discuss the applicability of the concept of Fano resonance, which, by definition, is realized in the two-oscillator model in the weak-coupling regime.



\section{Friedrich-Wintgen model}

In this section, we will briefly review the approach to describing BICs that arise from two interacting resonances via a radiation continuum due to appropriate tailoring of the structural parameters. The rigorous theory is presented in the paper by Friedrich and Wintgen \cite{Friedrich:1985}, and its simplified versions are described in a number of works. Following Ref. \cite{Seo:2020}, we consider the case when two optical resonances are coupled to the same radiation channel; the optical interaction between them is described by the non-Hermitian Hamiltonian without taking into account the nonradiative damping terms \cite{Hsu:2016,Koshelev:2019}:

\begin{equation}
H=\left( 
\begin{array}{cc}
E_{1} & k \\
k & E_{2}
\end{array} 
\right)
-i\left( 
\begin{array}{cc}
\gamma_{1} & e^{i\psi}\sqrt{\gamma_{1}\gamma_{2}} \\
e^{i\psi}\sqrt{\gamma_{1}\gamma_{2}} & \gamma_{2}
\end{array} 
\right)
=\left( 
\begin{array}{cc}
E_{1}-i\gamma_{1} & k-ie^{i\psi}\sqrt{\gamma_{1}\gamma_{2}} \\
k-ie^{i\psi}\sqrt{\gamma_{1}\gamma_{2}} & E_{2}-i\gamma_{2}
\end{array} 
\right)
\label{eq:ham}
\end{equation}

where, $E_{1}$ and $E_{2}$ are the resonance energy levels, $\gamma_{1}$  and  $\gamma_{2}$ are the corresponding radiative damping rates. Parameter $k$ denotes the internal (near-field) coupling strength between the modes and $\sqrt{\gamma_{1}\gamma_{2}}$  is the radiative coupling term, while $\psi$ is the phase difference between two modes. Therefore parameter $e^{i\psi}\sqrt{\gamma_{1}\gamma_{2}}$ represents the interference of radiating waves through far-field coupling \cite{Lee:2020}. The eigenvalues of Hamiltonian (\ref{eq:ham}) are determined by the secular equation and have the following form:

\begin{equation}
\widetilde{E}=\frac{E_{1}+E_{2}}{2}-i\frac{\gamma_{1}-\gamma_{2}}{2}\pm\frac{1}{2}\sqrt{\left(\frac{E_{1}-E_{2}}{2}-i\frac{\gamma_{1}-\gamma_{2}}{2}\right)^2+\left(k-ie^{i\psi}\sqrt{\gamma_{1}\gamma_{2}}\right)^2}
\label{eq:eigen}
\end{equation}

Equation (\ref{eq:eigen}) describes two avoided photonic bands in parametric space. Friedrich and Wintgen found a mathematical condition for one of the two eigenmodes to have a purely real eigenvalue. This condition has the form \cite{Friedrich:1985}:

\begin{equation}
\left({E_{1}-E_{2}}\right)p\sqrt{\gamma_{1}\gamma_{2}}=k\left(\gamma_{1}-\gamma_{2}\right)
\label{eq:cond}
\end{equation}

A parameter $p$ was also introduced, which describes the parity (+1 or -1) of the phase difference between the two modes (i.e., in-phase or out-of-phase), which corresponds to $\psi=\pi{l}$, where $l$ is an integer. Using Friedrich - Wintgen condition (\ref{eq:cond}) we obtain the following expressions for two complex energy levels:

\begin{equation}
\widetilde{E}_{A}=\frac{E_{1}+E_{2}}{2}-p\frac{k}{\sqrt{\gamma_{1}\gamma_{2}}}\frac{\gamma_{1}+\gamma_{2}}{2}
\label{eq:EA}
\end{equation}

\begin{equation}
\widetilde{E}_{B}=\frac{E_{1}+E_{2}}{2}+p\frac{k}{\sqrt{\gamma_{1}\gamma_{2}}}\frac{\gamma_{1}+\gamma_{2}}{2}-i(\gamma_{1}+\gamma_{2})
\label{eq:EB}
\end{equation}

Thus, when condition (\ref{eq:cond}) is satisfied, the first eigenmode $\widetilde{E}_{A}$ (\ref{eq:EA}) becomes purely real, corresponding to a BIC. The second eigenmode $\widetilde{E}_{B}$ becomes more radiative, the radiation losses of which are the sum of two initial modes, $\gamma_{B}=\gamma_{1}+\gamma_{2}$. It follows from equations (\ref{eq:EA},\ref{eq:EB}) that, depending on the sign of parity $p$ or coupling strength $k$, the BIC can be located on either the low or high frequency branch in parametric space.

\section{Calculation methods}

To study the transformation of quasi-BIC when the structure of the resonator changes from a dielectric cylinder to a dielectric ring, we used numerical calculations, which provide key information about the optical spectrum with resonant frequencies of eigenmodes and mode $Q$-factor. Structures with a fixed homogeneous dielectric permittivity $\epsilon_{1}=80$  and zero attenuation were considered. The environment was vacuum  $\epsilon_{2}=1$. We have performed systematic calculations of the scattering cross-section (SCS) of a finite cylinder and a set of rings with an inner radius $R_{in}$, an outer radius $R_{out}$ and a thickness $L$. The ratio of the radii was varied in the wide  range, and the ${R_{out}}/{L}$  aspect ratio for each $R_{in}$ value was varied in such a range to observe a certain quasi-BIC formed by the anticrossing of the two resonant modes $\mbox{TE}_{1,1,0}$ and $\mbox{TM}_{1,1,1}$. The results of the study of such a quasi-BIC in a cylinder are presented in Ref. \cite{Bogdanov:2019}, where the standard nomenclature \cite{Zhang:2008} of modes of a cylindrical resonator $\mbox{TE}_{n,k,p}$ and $\mbox{TM}_{n,k,p}$ is used. In these notations, $n, k, p$ are the indices denoting the azimuthal, radial, and axial indices, respectively. Only the modes with the same azimuthal index n could interact and we consider the case $n$ = 1. In our calculations, the magnetic field of the incident wave is polarized along the axis of rotation of the cylinder or ring (TE polarization). All the computations of SCS were performed in the frequency domain using the commercial software COMSOL. The calculations of the spectra of the dielectric cylinder were carried out on the basis of the Mie theory \cite{Bohren:1983}. 

Mie scattering by high-contrast dielectric bodies of revolution results in an infinite series of Fano resonances where each resonance can be described by the Fano formula. The conclusion was made for the case of dielectric cylinder \cite{Rybin:2013,Rybin:2015} and later for dielectric sphere \cite{Tribelsky:2016} and core-shell sphere \cite{Arruda:2018}. The Fano resonance is a consequence of the interference of two waves, which in the spectra have significantly different half-widths \cite{Fano:1961,Limonov:2017}. The shape of the resulting contour of a narrower line depends on the phase difference $\delta$ between them. The frequency and width of the Fano line is determined by the well-known formula \cite{Limonov:2017}:

\begin{equation}
S(\omega)=D^2\frac{(q+\Omega)^2}{1+{\Omega}^2}
\label{eq:Fano}
\end{equation}
where $q = \cot\delta$ is the Fano asymmetry parameter, $\delta$ is the phase difference between a discrete state and a continuum, $D^2=4\sin^2\delta$, $\Omega=(\omega-\omega_{0})/(\gamma_{0}/2)$, is the dimension-less frequency, and $\gamma_{0}$ and $\omega_{0}$ are the width and frequency of the narrow line, respectively. All numerically obtained spectra were fitted using formula (\ref{eq:Fano}), as a result of which the frequencies, Fano parameters $q$, and $Q$-factors of two interfering modes were determined.

%
\begin{figure*}
  \includegraphics[width=0.60\textwidth]{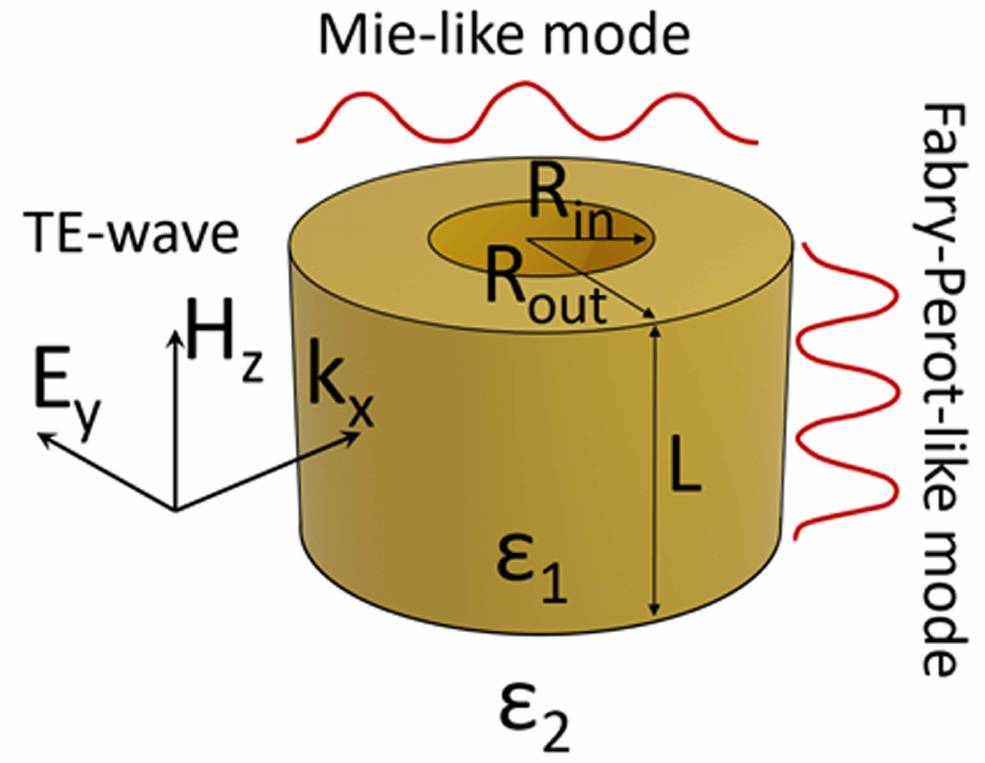}
  \caption
{
TE- and TM-polarized waves incident on a dielectric ring resonator with permittivity $\epsilon_{1}=80$, inner radius $R_{in}$, outer radius $R_{out}$ and thickness $L$. The ring is placed in the vacuum, $\epsilon_{2}=1$.
}
  \label{fig:fig1}       
\end{figure*}
%

When changing the aspect ratio $R_{out}/L$ (for example, by changing the length $L$ with a fixed outer radius $R_{out}$), the frequency of the resonant Fabry-Perot modes changes significantly, while the resonant Mie modes weakly (\ref{fig:fig1}). Therefore, the Fabry-Perot -type modes and Mie-type modes must intersect at certain parameters of the cylinder and the ring ($\epsilon, R_{out}/L, R_{in}/R_{out}$). In the case of their interaction, the avoided crossing regime will be observed, however, if there is no interaction or it is insufficient, the branches will intersect in the parametric space.Continuing the work presented in Ref. \cite{Bogdanov:2019}, we studied the behavior of two modes that have the symmetry $\mbox{TE}_{1,1,0}$ (slow-varying Mie-type mode) and $\mbox{TM}_{1,1,1}$ (rapidly-varying Fabry-Perot-type mode). 

\section{Results and discussion}

%
\begin{figure*}
  \includegraphics[width=0.95\textwidth]{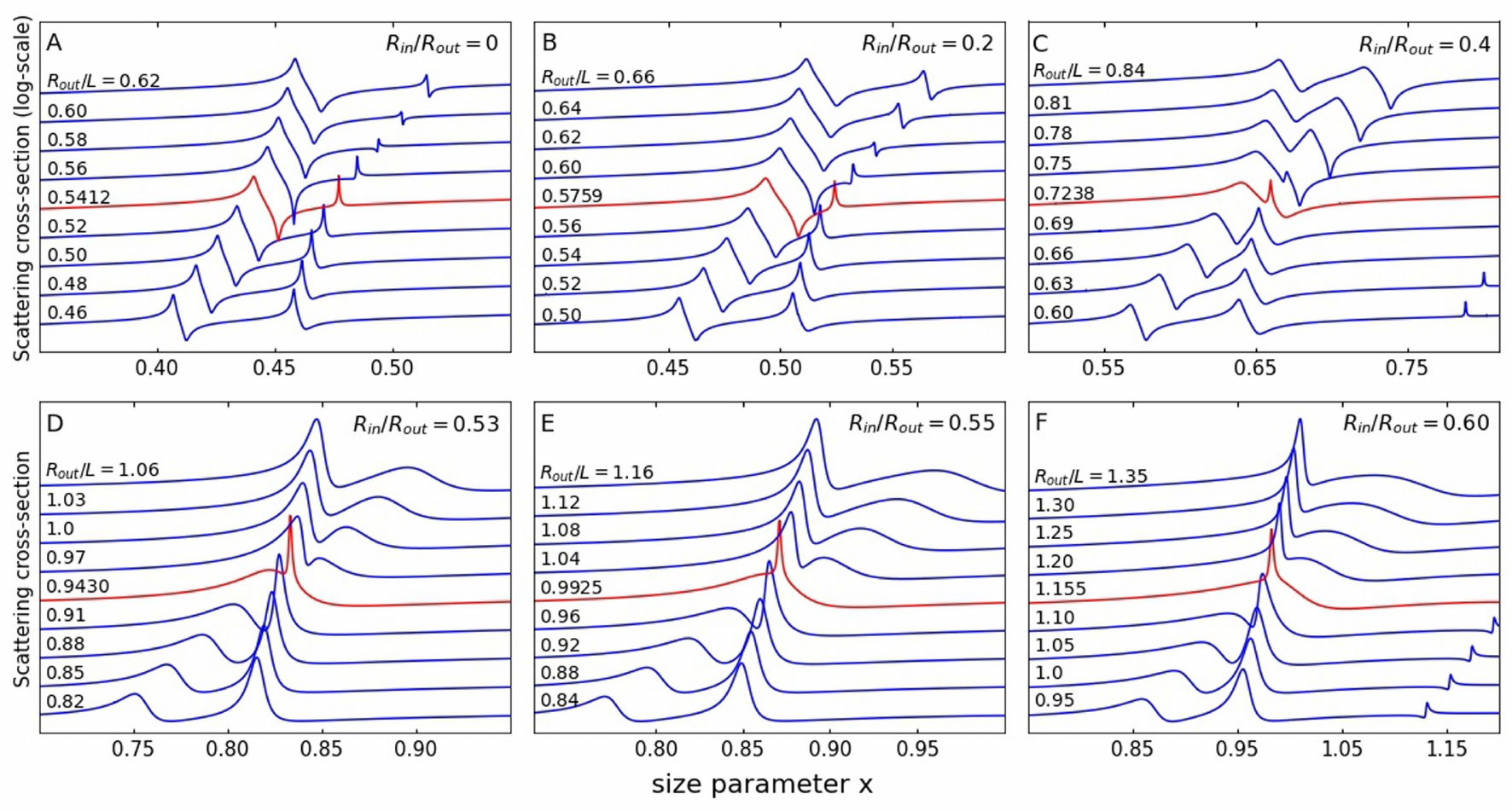}
  \caption
{
Spectra of the normalized SCS (for harmonic $n=1$) for of cylindrical (A) and ring (B-F) dielectric resonators as a function of their aspect ratio $R_{out}/L$ in the regions of avoided crossing (A-C) or crossing (D-F) regimes between the modes $\mbox{TE}_{1,1,0}$ and $\mbox{TM}_{1,1,1}$. The spectra marked in red correspond to quasi-BIC. Normalized size parameter (frequency) $x=R_{out}\omega/c=2\pi{R_{out}}/\lambda$  TE-polarized incident wave. The dielectric permittivity of all structures is $\epsilon_{1}=80$. The structures are placed in the vacuum, $\epsilon_{2}=1$.
}
  \label{fig:fig2}       
\end{figure*}
%

This section presents the results of calculations of SCS, their treatment and interpretation. To obtain a complete picture of light scattering in dielectric ring resonators, we repeated the calculations described in Ref. \cite{Bogdanov:2019} for a finite dielectric cylinder and used these results as a reference point. Further, calculations were made for a large number of ring resonators with different values of the normalized inner diameter in the range $0\le{R_{in}}/R_{out}\le0.6$. For each fixed value of $R_{in}/R_{out}$, the SCS of the dielectric ring resonator were calculated as a function of its aspect ratio $R_{out}/L$ upon excitation by a plane TE-polarized wave (see Figure \ref{fig:fig2} for selected $R_{in}/R_{out}$). As in Refs. \cite{Rybin:2017} and \cite{Bogdanov:2019}, the dielectric permittivity was $\epsilon_{1}=80$. With this value, firstly, the resonance scattering spectra have narrow peaks that are convenient for treatment and interpretation and, secondly, this value corresponds to the permittivity of water in the microwave frequency region at room temperature, which in the future may allow an experiment to be carried out and its results compared with the presented data. For generality, when demonstrating the results we use the normalized size parameter $x=kR_{out}$ being a product of the wavenumber $k$ and outer radius $R_{out}$.

%
\begin{figure*}
  \includegraphics[width=0.60\textwidth]{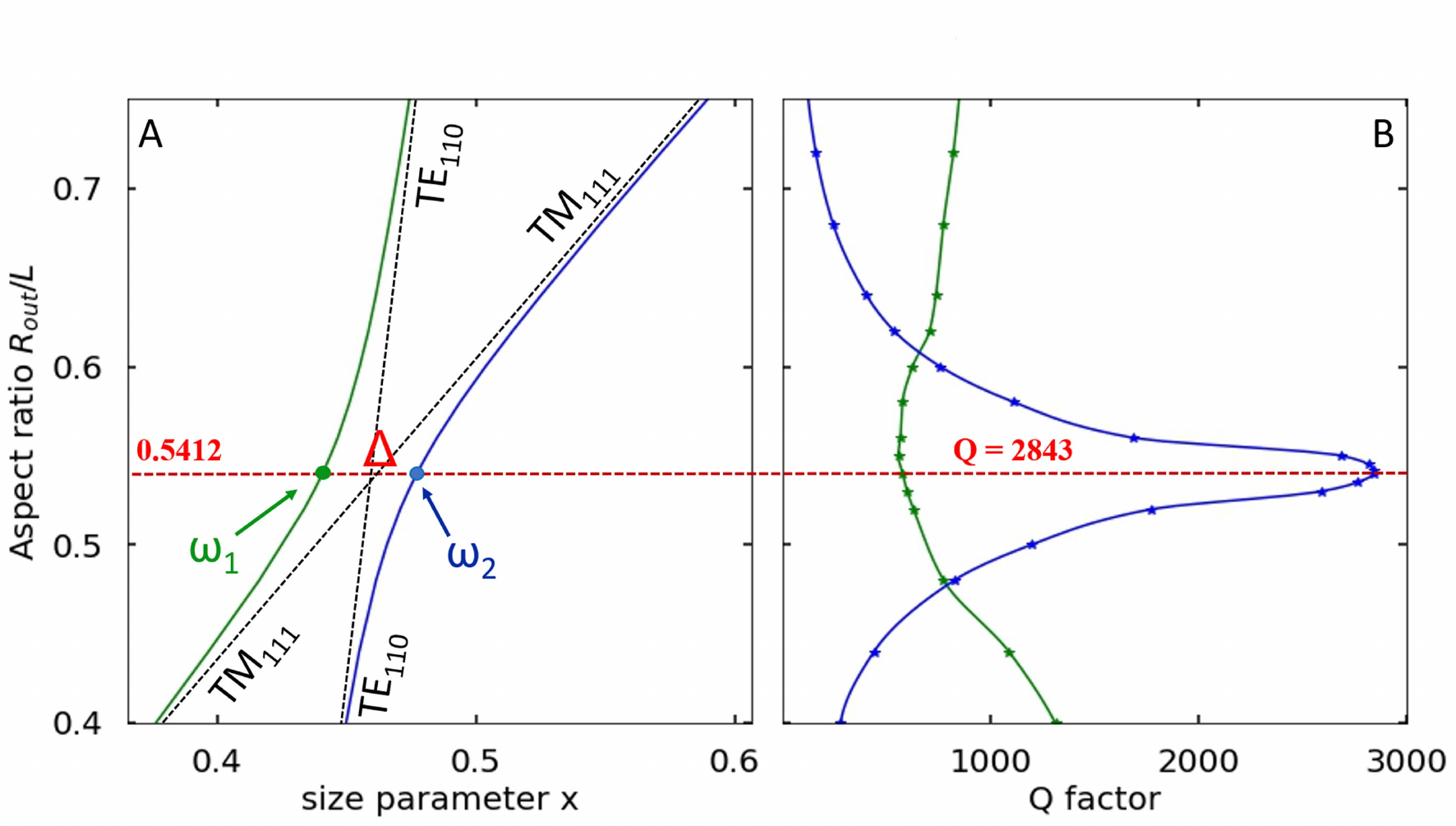}
  \caption
{
(A) $\mbox{TE}_{1,1,0}$ (slow-varying Mie-type mode) and $\mbox{TM}_{1,1,1}$ (rapidly-varying Fabry-Perot-type mode) in finite cylindrical dielectric resonator. The parameter $\Delta$ corresponds to the frequency difference between the high-frequency and low-frequency branches (Rabi splitting $ \Delta=\omega_{2}-\omega_{1}$) at the aspect ratio for the quasi-BIC $(R_{out}/L=0.5412)$. (B) $Q$-factor evolution for the high-frequency branch (blue curve, quasi - BIC) and the low-frequency branch (green curve). $\epsilon_{1}=80$, $\epsilon_{2}=1$.
}
  \label{fig:fig3}       
\end{figure*}
%

\begin{figure*}
  \includegraphics[width=0.95\textwidth]{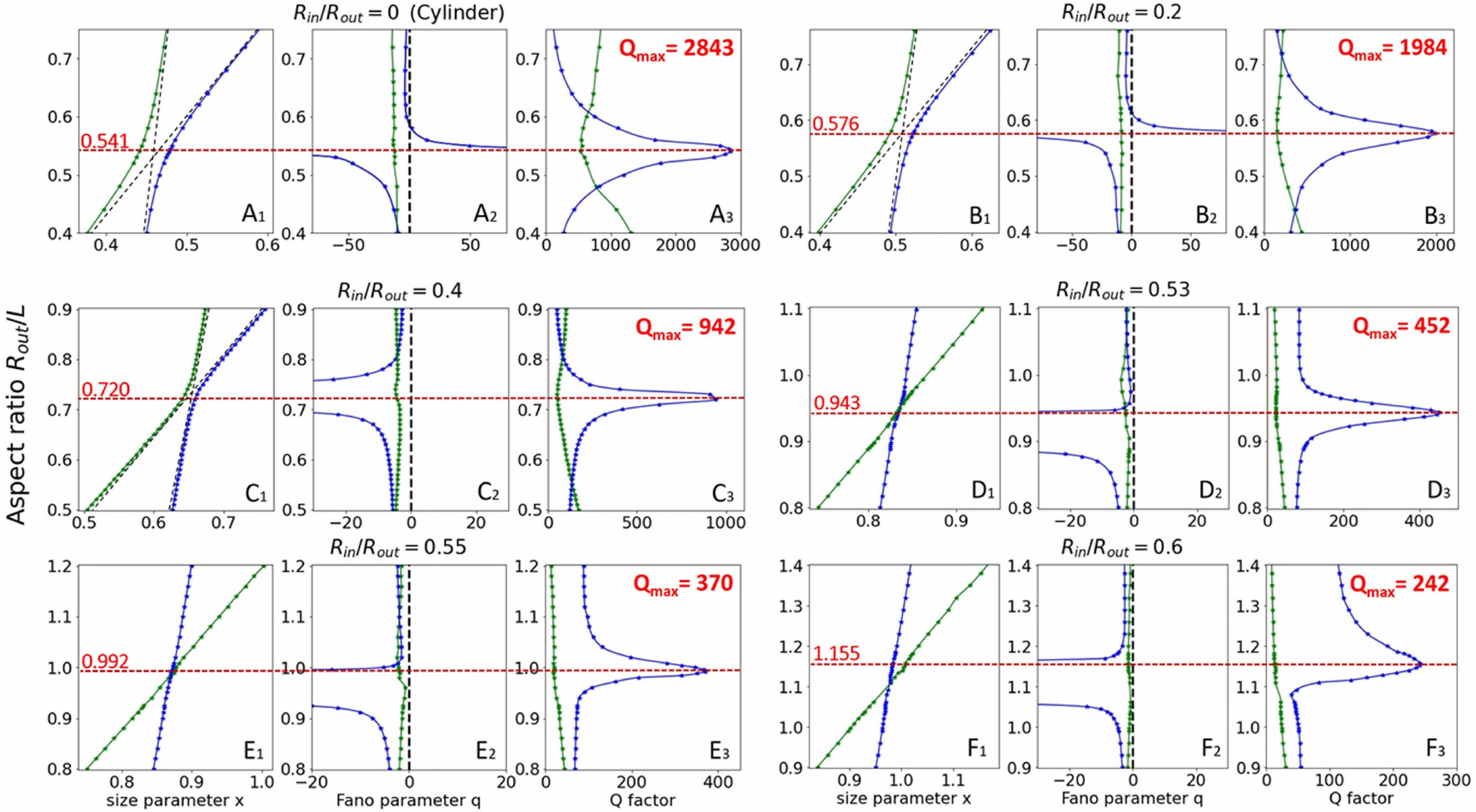}
  \caption
{
Results of treatment the SCS spectra (for harmonic $m=1$) using the Fano formula (\ref{eq:Fano}) for the low-frequency and high-frequency modes. (A1-F1) Frequencies of the $\mbox{TE}_{1,1,0}$ (Mie-type mode) and $\mbox{TM}_{1,1,1}$ (Fabry-Perot-type mode). (A2-F2) Evolution of the Fano asymmetry parameter $q$. (A3-F3) Evolution of the quality-factor $Q$. Panels with index (A) correspond to the cylinder, and those with indexes (B-F) correspond to rings with the ratio ${{R_{in}}/R_{out} = 0.2, 0.4, 0.53, 0.55, 0.6}$ respectively.
}
  \label{fig:fig4}       
\end{figure*}

To fit the spectra and determine the parameters of the resonance lines, we used the Fano formula (\ref{eq:Fano}). For clarity, let us consider separately different regions in the parametric space: first, two regions of weak-coupling above and below the region of avoided crossing, where the corresponding branches have a linear dependence on the parameter $R_{out}/L$, and, second, a region of strong-coupling, where the avoided crossing is observed. In areas of weak interaction, asymmetric contours of resonance scattering lines are observed (Figure \ref{fig:fig2}), associated with the interference of each of the two resonant modes ($\mbox{TE}_{1,1,0}$ and $\mbox{TM}_{1,1,1}$) with broadband background scattering from the dielectric object as a whole \cite{Rybin:2013,Rybin:2015,Tribelsky:2016}. As has been shown earlier, these spectra exhibit Fano resonances, a phenomenon related to the case of two weakly coupled oscillators \cite{Limonov:2017}. As a result of fitting, the frequencies, Fano parameters $q$, and $Q$-factors of the pairs of lines, presented in Figures \ref{fig:fig3} and \ref{fig:fig4}, were determined with high accuracy. 

In the quasi-BIC region of the avoided crossing, we have the interference of three waves in the far-field, two waves are determined by the resonant modes of a cylinder or ring ($\mbox{TE}_{1,1,0}$ and $\mbox{TM}_{1,1,1}$), and the third wave is associated with a nonresonant background. Such complicated interference is not described by the classical Fano formula; therefore, the statement that a collapse of the Fano resonance occurs in the quasi-BIC region \cite{Azzam:2018,Pilipchuk:2020,Bonod:2020} is correct. Accordingly, one should not expect a perfect fit according to the Fano formula where the condition of its applicability is not met. Nevertheless, the shape of the resonance lines in the spectra corresponding to the avoided crossing region was successfully approximated by the Fano formula in a significant region, excluding the extremely narrow quasi-BIC zone, where the shape of the high-frequency line corresponded to the symmetric Lorentzian, the Fano parameter $q$ tends to infinity, and the $Q$-factor reached its maximum value.

%
%

Quasi-BIC due to the avoided crossing of the Mie-type mode $\mbox{TE}_{1,1,0}$ and Fabry-Perot-type mode $\mbox{TM}_{1,1,1}$, at   is observed in a cylinder with a vertical cross-section close to square $R_{out}/L=0.54$ at a size parameter of $x = 0.48$ (Figure \ref{fig:fig2}A). To find the avoided crossing region of interest to us in the spectra of the ring, it was necessary to significantly shift the search regions both in the frequency scale and in the $R_{out}/L$ scale [Figures \ref{fig:fig2}(B-F)]. With an increase in the normalized inner radius $R_{in}/R_{out}$, the avoided crossing region in parametric space shifted towards thinner rings, i.e. parameter $R_{out}/L$  increased. With relatively small internal holes ($R_{in}/R_{out}\le{0.4}$) the avoided crossing regime is still observed. For small values of the aspect ratio $R_{out}/L$, the Fabry-Perot resonance determines the lower-frequency mode, and the Mie resonance determines the higher-frequency mode; above the quasi-BIC region, the resonances are reversed. In this case, outside the avoided crossing region, both resonances demonstrate a linear dependence of the frequency on the parameter $R_{out}/L$. Outside the avoided crossing region, the frequency shifts of both Mie-type and Fabry-Perot-type modes are described by linear relations. Nevertheless, an important tendency is observed: with an increase in the size of the inner hole $R_{in}/R_{out}$, the distance between the low-frequency and high-frequency resonances at the quasi-BIC frequency (Rabi splitting \cite{Khitrova:2006}  $\Delta=\omega_{2}-\omega_{1}$, Figure \ref{fig:fig3}) decreases, Figure \ref{fig:fig4}.   

The key effect is observed in the region of the $R_{in}/R_{out}={0.53}$ parameter, when the lines in the spectra corresponding to the $\mbox{TE}_{1,1,0}$ and $\mbox{TM}_{1,1,1}$ resonances intersect and the avoided crossing effect disappears completely. The spectra clearly show a simple superposition of a narrow $\mbox{TE}_{1,1,0}$ line on a broad $\mbox{TM}_{1,1,1}$ line without specific signs of interference (Figure \ref{fig:fig2}D). It is important to note that in this case, the quasi-BIC regime is still preserved, although the maximum value of the quality factor $Q = 452$ decreased significantly compared to the value $Q = 2843$ in the cylinder. We confidently call the observed effect of quasi-BIC, since the usual resonance does not have a pronounced maximum in the parametric space, and the quasi-BIC should have such a maximum \cite{Rybin-Nat:2017} that is actually observed in the spectra of the ring at $R_{in}/R_{out}={0.53}$ (Figure \ref{fig:fig4}D3) and moreover, with even larger values of the hole $R_{in}/R_{out}={0.55}$ and $R_{in}/R_{out}={0.60}$ (Figures \ref{fig:fig4}E3 and \ref{fig:fig4}F3). 

Thus, we observe a continuous transition from the avoided crossing of two branches (strong-coupling regime) to their intersection with the preservation of the quasi-BIC (weak-coupling regime). It is a degenerate point in parametric space ($R_{in}/R_{out}={0.53}$, Figure \ref{fig:fig4}D1) where two dispersion curves intersect with straight lines. In the region of $R_{in}/R_{out}\le{0.53}$, the quasi-BIC is associated exclusively with the $\mbox{TE}_{1,1,0}$ mode of the Mie type (Figures \ref{fig:fig4}D3, \ref{fig:fig4}E3, \ref{fig:fig4}F3), while in the cylinder (Figures \ref{fig:fig3}, \ref{fig:fig4}A3) and ring with small holes (Figures \ref{fig:fig4}B3, \ref{fig:fig4}C3), the quasi-BIC was determined by the strong interaction of the Mie-type $\mbox{TE}_{1,1,0}$ and Fabry-Perot-type $\mbox{TM}_{1,1,1}$ resonances.

%
\begin{figure*}
  \includegraphics[width=0.95\textwidth]{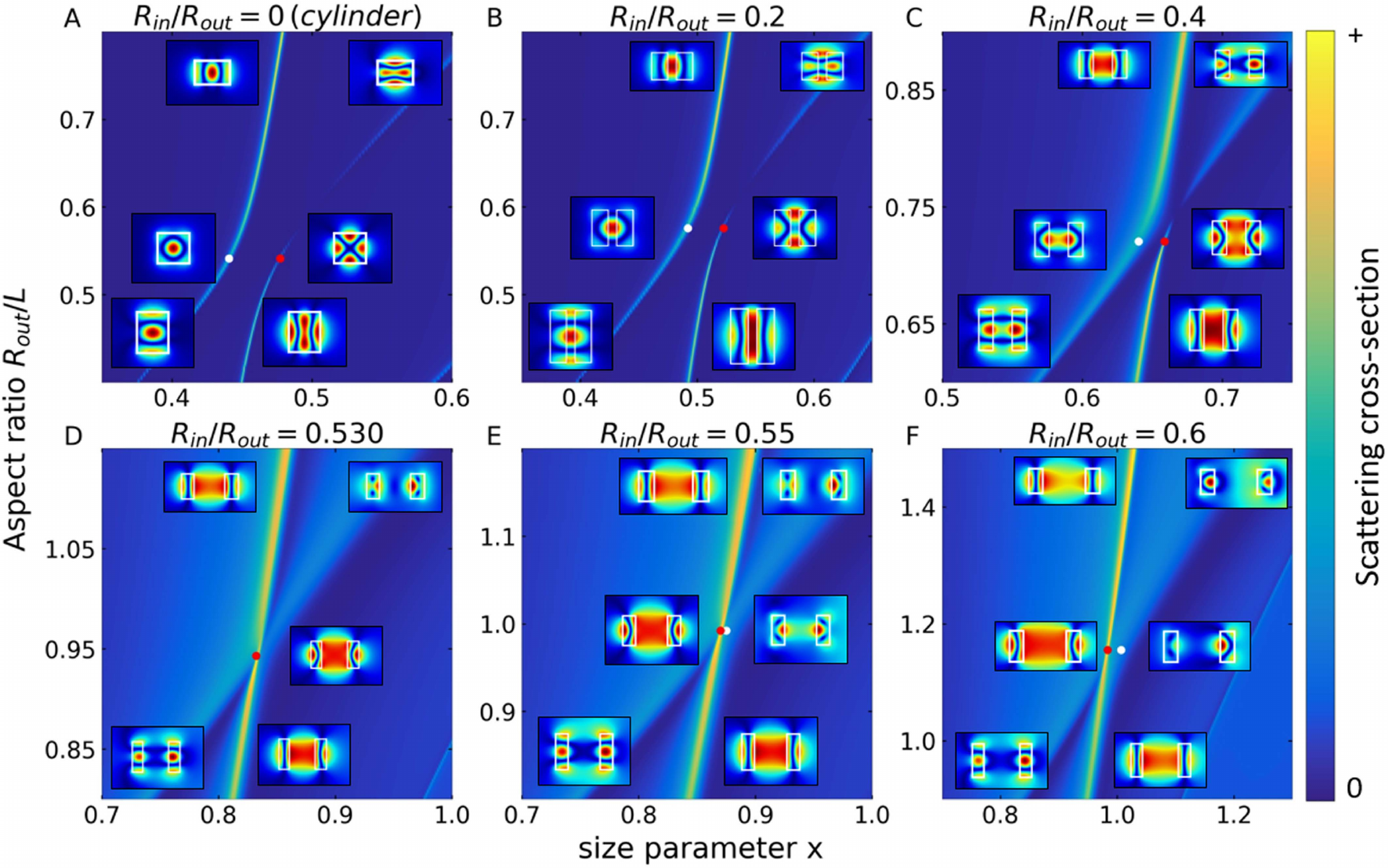}
  \caption
{
Dependencies of the SCS $(m=1)$ of cylindrical (A) and ring (B-F) dielectric resonators on the aspect ratio $R_{out}/L$ and normalized size parameter (frequency) $x=R_{out}\omega/c=2\pi{R_{out}}/\lambda$ for TE-polarized incident waves. The calculations are carried out with the $R_{out}/L$ step of $0.02$. Inserts: The calculated field patterns (total electric field amplitude $\vert{E}\vert$  in the $(x,z)$ cross sections (resonator side view) for the high-frequency and low-frequency branches of the interacting pair of modes $\mbox{TE}_{1,1,0}$ and $\mbox{TM}_{1,1,1}$. The field distributions are shown for three $R_{out}/L$ values - two extreme ones, shown on each of the panels and the $R_{out}/L$ value corresponding to the quasi-BIC. Quasi-BIC points are marked with red dots, and points corresponding to the same $R_{out}/L$ value on a different branch are marked with white dots. 
}
  \label{fig:fig5}       
\end{figure*}
%

Figure \ref{fig:fig5} shows a color pattern of the SCS in the size parameter and aspect ratio axes. It is clearly seen that with an increase in the size of the inner hole, the lines in the spectra are significantly broadened, especially for the $\mbox{TM}_{1,1,1}$ Fabry-Perot-type mode. The same conclusion follows from Figure \ref{fig:fig2}. The difference in the behavior of Mie-type and Fabry-Perot-type modes is qualitatively clear. Resonant Fabry-Perot modes are defined by the top and bottom faces of the cylinder or ring. With a decrease in the effective area of these faces in the ring, with an increase in $R_{in}$, the $Q$-factor of the vertical resonances inevitably decreases. For Mie resonances, the situation is different, since the area of the side wall does not depend on the diameter of the inner hole $R_{in}$. As follows from the calculations (Figure \ref{fig:fig5}), the electric field of the Mie-type mode is concentrated in the inner air cylinder over the entire range of parameters. It is important to emphasize that in the avoided crossing region (especially for cylinder and ring with $R_{in}/R_{out}={0.2}$), the electric field amplitude $\vert{E}\vert$ of the quasi-BIC is determined by the combination of the Fabry-Perot and Mie resonance fields, but in the crossing regime the field of the quasi-BIC (red dots in Figure \ref{fig:fig5}) is completely determined by Mie-type resonance.

%
\begin{figure*}
  \includegraphics[width=0.95\textwidth]{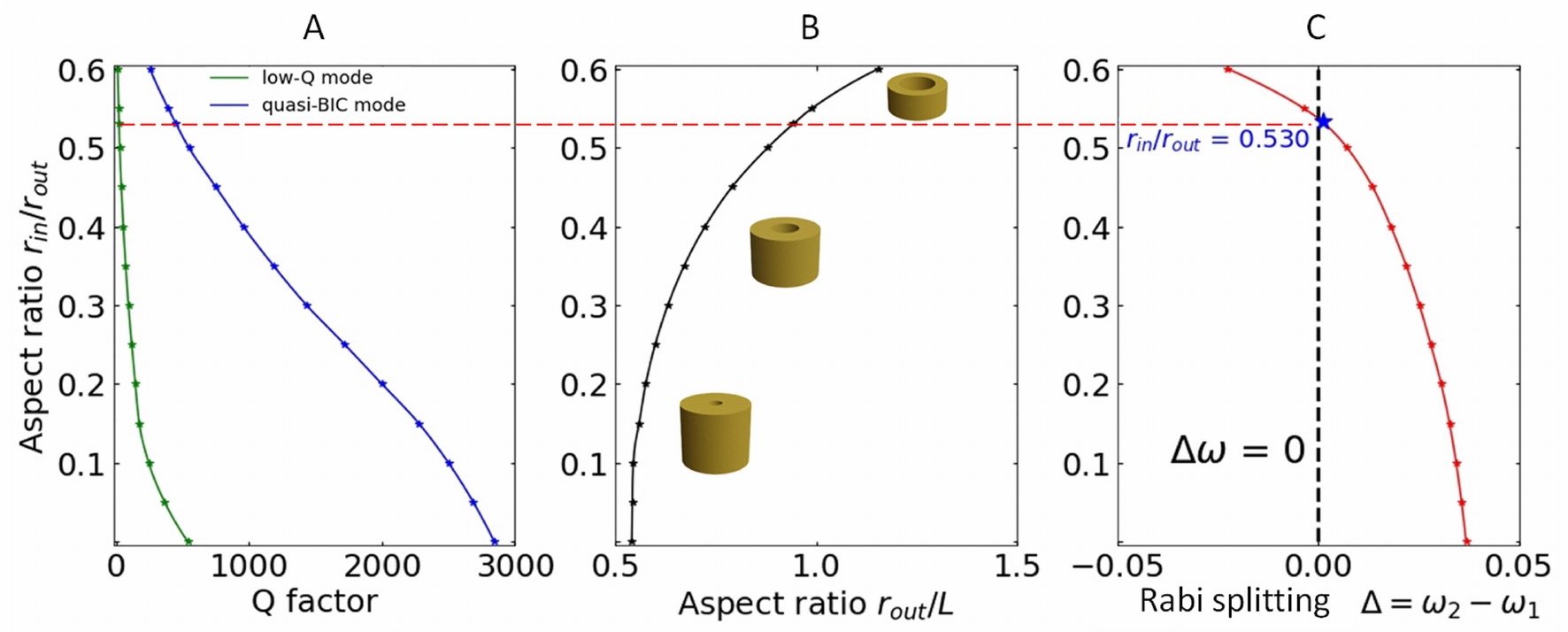}
  \caption
{
(A) $Q$-factor evolution for branches with the quasi-BIC (blue curve) and without quasi-BIC (green curve). (B) Transformation of the shape of the ring (ratios $L$, $R_{out}$, $R_{in}$) with the quasi-BIC for a pair of modes $\mbox{TE}_{1,1,0}$ and $\mbox{TM}_{1,1,1}$. (C) Dependence of the Rabi  splitting  ($\Delta=\omega_{2}-\omega_{1}$) between the branch with the quasi-BIC and a branch without the quasi-BIC on the normalized radius of the inner hole $R_{in}/R_{out}$.  
}
  \label{fig:fig6}       
\end{figure*}
%

Figure \ref{fig:fig6} summarizes the results of this work. Smooth dependences of all functions leave no doubt that the quasi-BIC is preserved in the case of crossing the branches and further, with an increase in $R_{in}/R_{out}$, when the frequency of the Mi-type quasi-BIC (red dots in Figure \ref{fig:fig5}) turns out to be lower than the Fabry-Perot resonance frequency (white dots) for the same $R_{out}/L$ value.
We note that the results obtained using the Fano formula contain key information about the quasi-BIC. There is a direct relationship between the Friedrich-Wintgen quasi-BICs and Fano resonances, since these two phenomena are associated with the same physical effect - wave interference. Note the difference between quasi-BIC and true theoretical BIC, which is a dark non- radiative state with infinite $Q$ and, accordingly, should manifest itself in the scattering spectra as a dip exactly to zero. In contrast to these properties of a true BIC, the quasi- BIC has two singular points in spectra, characterized by the Fano parameter $q$. The position of the quasi- BIC is determined by the maximum $Q$-factor of the line and corresponds to the Fano parameter $q\to\infty$. At this value, the Fano lineshape becomes a symmetric Lorentzian function and the resonance does not couple to the continuum of states. Accordingly, the minimum of scattering does not coincide with the position of the quasi- BIC and corresponds to the anti-resonance in the continuum spectrum ($q=0$), and appears as a true zero in the spectrum. However, this minimum may not reach zero if more continuum states become involved in the scattering process \cite{Limonov:2017}. 
As follows from expressions (\ref{eq:EA}) and (\ref{eq:EB}), the magnitude of the splitting of $\mbox{TE}_{1,1,0}$ and $\mbox{TM}_{1,1,1}$ branches has the form:

\begin{equation}
\Delta\sim{\widetilde{E}_{B}-\widetilde{E}_{A}}=2p\frac{k}{\sqrt{\gamma_{1}\gamma_{2}}}\frac{\gamma_{1}+\gamma_{2}}{2}-i(\gamma_{1}+\gamma_{2})
\label{eq:Delta}
\end{equation}

As follows from Figure \ref{fig:fig6}C, the difference $\Delta\sim{\widetilde{E}_{B}-\widetilde{E}_{A}}$ in the avoided crossing region changes monotonically, which indicates a monotonic change in the internal coupling strength $\kappa$ at a constant parity parameter $p$ which can only change in a jump from +1 to  1. Consequently, the intersection of the $\mbox{TE}_{1,1,0}$ and $\mbox{TM}_{1,1,1}$ branches, that is, the relation $Re(\widetilde{E}_{B})=Re(\widetilde{E}_{A})$ indicates the proximity to zero of the parameter $\kappa$. In this case, expressions for resonant energy levels can be written in a simpler form:

\begin{equation}
\widetilde{E}_{A}=\frac{E_{1}+E_{2}}{2}
\label{eq:EA_simple}
\end{equation}

\begin{equation}
\widetilde{E}_{B}=\frac{E_{1}+E_{2}}{2}-i(\gamma_{1}+\gamma_{2})
\label{eq:EB_simple}
\end{equation}

Thus, the Friedrich - Wintgen model predicts that when crossing, two lines with significantly different half-widths ($\gamma_{A}=0$  and $\gamma_{B}=\gamma_{1}+\gamma_{2}$) will be observed in the spectra, which corresponds to the result that was obtained in our calculations. The eigenmode $\widetilde{E}_{A}$ continues to be purely real and preserves the quasi-BIC regime.

It should be noted that individually regimes of strong coupling and weak coupling of two photonic modes were previously studied in various dielectric resonators. In particular, the formation of long-lived states with high Q-factor near avoided resonance crossings due to strong external coupling was studied in quasi-two-dimensional dielectric cavities with rectangular, elliptical, and stadium-shaped cross section \cite{Wiersig:2006}. External weak coupling with crossing resonant frequencies also leads to a significant increase in Q-factor, as demonstrated for the case with three dielectric nanorods \cite{Song:2010}. This regime was achieved by tuning of the rod spacing and led to an increase in the lifetime over an order of magnitude.

\section{Conclusion}

We have revealed that high-$Q$ quasi-BIC exists not only in dielectric cylindrical resonators, but also in dielectric ring resonators with high robustness over a wide range of internal hole sizes. The behavior of the Mie-type $\mbox{TE}_{1,1,0}$ and Fabry-Perot-type $\mbox{TM}_{1,1,1}$ resonances in structures with a dielectric permittivity $\epsilon_{1}=80$ was investigated numerically. Destructive interference of anti-phase waves in the far zone leads to a quasi-BIC, and constructive interference of in-phase waves leads to a decrease in the $Q$ factor. In the quasi-BIC region of the avoided crossing, there is interference of three waves in the far-field zone, two waves are determined by the resonant modes of a cylinder or ring ($\mbox{TE}_{1,1,0}$ and $\mbox{TM}_{1,1,1}$ as an example), and the third wave is associated with a nonresonant background. Such a complex interference is not described in the classical Fano model and this effect is associated with the collapse of the Fano resonance. At $R_{in}/R_{out}\sim{0.53}$, we found a crossover from the region of avoided crossing to the region of intersection of branches in the parametric space. The crossover occurs due to a monotonic decrease to the minimum values of the internal coupling strength $\kappa$ at a constant parity parameter $p$. In the region of intersection, the quasi-BIC is preserved and is observed in the spectra exclusively on the Mie-type $\mbox{TE}_{1,1,0}$ line which is much narrower than the $\mbox{TM}_{1,1,1}$ line. With a change in the aspect ratio $R_{out}/L$, the $\mbox{TE}_{1,1,0}$ line demonstrates the maximum $Q$ factor associated with quasi-BIC when it is at the top of the $\mbox{TM}_{1,1,1}$ broad band. 

In this work, we demonstrate in detail how to perform the fine tuning of structural parameters ($\epsilon_{1}$, $\epsilon_{2}$, $L$,$R_{in}$, $R_{out}$) for designing a dielectric ring with the maximum $Q$ factor in a certain spectral range due to the quasi-BIC regime. Note that when the size of the inner hole changes from zero in a fairly wide range, the $Q$-factor of the quasi-BIC changes insignificantly, demonstrating high robustness. 

In this context, we would like to draw attention to chalcogenide phase-change alloys Ge-Sb-Te. In particular, the femtosecond laser-induced switching of infrared nanoantenna resonances using a $\mbox{Ge}_{3}\mbox{Sb}_{3}\mbox{Te}_{6}$ thin film was reported \cite{Michel:2014}. Another Ge-Sb-Te alloy, namely $\mbox{Ge}_{2}\mbox{Sb}_{2}\mbox{Te}_{5}$, is also being actively investigated \cite{Wang:2016}. In particular, $\mbox{Ge}_{2}\mbox{Sb}_{2}\mbox{Te}_{5}$ metasurfaces were created using U-shaped antennas \cite{Choi:2019} or a complex nanostructured cruciform shape developed using a genetic algorithm optimizer in combination with an efficient full-wave electromagnetic solver \cite{Pogrebnyakov:2018}. We believe that the use of easily manufactured cylindrical or annular metasurface elements formed from Ge-Sb-Te in the quasi-BIC mode will significantly improve the functional properties of various devices.

\begin{acknowledgement}
  The authors are grateful to P.A. Belov for discussing the results.
\end{acknowledgement}

\begin{funding}
  This work was funded by the Russian Science Foundation (Project 20-12-00272). 
  
http://dx.doi.org/10.13039/501100006769,"Russian Science Foundation"
\end{funding}

\end{document}